\RequirePackage{snapshot}
\documentclass{IEEEtran4PSCC}
\ifCLASSINFOpdf
   \usepackage[pdftex]{graphicx}
\else
   \usepackage[dvips]{graphicx}
\fi
\usepackage[cmex10]{amsmath}
\ifCLASSOPTIONcompsoc
  \usepackage[caption=false,font=normalsize,labelfont=sf,textfont=sf]{subfig}
\else
  \usepackage[caption=false,font=footnotesize]{subfig}
\fi
 \usepackage{url}

\usepackage{booktabs}
\usepackage[referable]{threeparttablex}
\usepackage{multirow}
\usepackage[noabbrev]{cleveref}

\makeatletter
\let\old@ps@headings\ps@headings
\let\old@ps@IEEEtitlepagestyle\ps@IEEEtitlepagestyle
\def\psccfooter#1{%
    \def\ps@headings{%
        \old@ps@headings%
        \def\@oddfoot{\strut\hfill#1\hfill\strut}%
        \def\@evenfoot{\strut\hfill#1\hfill\strut}%
    }%
    \def\ps@IEEEtitlepagestyle{%
        \old@ps@IEEEtitlepagestyle%
        \def\@oddfoot{\strut\hfill#1\hfill\strut}%
        \def\@evenfoot{\strut\hfill#1\hfill\strut}%
    }%
    \ps@headings%
}
\makeatother

\psccfooter{%
        \parbox{\textwidth}{\hrulefill \\ \small{21st Power Systems Computation Conference} \hfill \begin{minipage}{0.2\textwidth}\centering \vspace*{4pt} \includegraphics[scale=0.06]{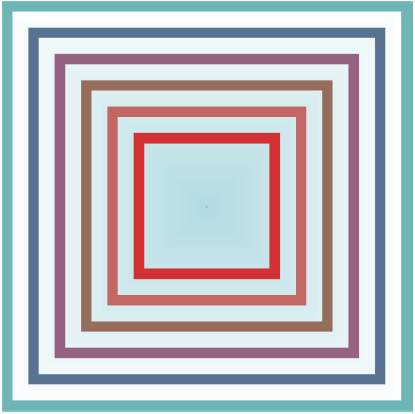}\\\small{PSCC 2020} \end{minipage} \hfill \small{Porto, Portugal --- June 29 -- July 3, 2020}}%
}

\newcommand{\figWidth}{0.9\linewidth}

\begin{document}
%
\title{Mitigating the impact of distributed PV in a low-voltage grid using electricity tariffs}

\author{
\IEEEauthorblockN{Jordan Holweger, Lionel Bloch, Christophe Ballif, Nicolas Wyrsch}
\IEEEauthorblockA{Photovoltaics and thin film electronics laboratory (PV-LAB) \\
\'Ecole Polytechnique F\'ed\'erale de Lausanne (EPFL), Institute of Microengineering (IMT)\\
Neuch\^atel, Switzerland\\
\{jordan.holweger, lionel.bloch, christophe.ballif, nicolas.wyrsch\}@epfl.ch}
}


\maketitle



\begin{abstract}
A high share of distributed photovoltaic (PV) generation in low-voltage networks may lead to over-voltage, and line/transformer overloading. To mitigate these issues, we investigate how advanced electricity tariffs could ensure safe grid operation while enabling building owners to recover their investment in a PV and storage system. We show that dynamic volumetric electricity prices trigger economic opportunities for large investments in PV and battery capacity but lead to more pressure on the grid while capacity and block rate tariffs mitigate over-voltage and decrease line loading issues. However, block rate tariffs significantly decrease the optimal PV installation size.
\end{abstract}


\thanksto{ Swiss Centre for Competence in Energy Research on the Future Swiss Electrical Infrastructure (SCCER-FURIES)}

\vspace{-0.8cm}
\section{Introduction}
Significant investments in photovoltaic (PV) systems are expected in the coming years due to increasing economic interest and to their contribution to clean energy generation. Reaching the full PV potential in urban environments could lead a system in which a significant fraction of the energy should be curtailed in order to cope with the network operating constraints. Centralized control of PV curtailment by the distribution system operator is a promising solution but could raise concerns about intrusiveness and discourage (potential) prosumers. An alternative approach is to propose novel electricity pricing mechanisms that help to mitigate the impact of distributed storage and PV systems on the grid while allowing building owners to make profitable investments. The purpose of this work is to evaluate which advanced electricity tariff creates the best trade-off between these two objectives. 

The impact of the high penetration of distributed renewable energy sources is investigated in \cite{Tonkoski2012} where it is shown how the network topology and the way distributed PV is geographically located along the feeder have a strong impact on the voltage profile.  Considering the deployment of net-zero-energy buildings, Arboleya et al. \cite{Arboleya2017} study the impact of various PV penetration levels on the voltage profile of a low-voltage grid. The authors of this study assume that each building minimize its energy bill, i.e. no coordination between building is present, and argue that the network will have to face such uncoordinated operation for years to come. Conversely, Hidalgo-Rodriguez et al. \cite{Hidalgo-Rodriguez2018} study two micro-grid coordinated control strategies and one uncoordinated control strategy to achieve network balancing. Further in this direction, Haschemipour et al. \cite{Hashemipour2018} propose a control strategy for PV and battery systems to achieve voltage regulation. The authors, however, do not consider the profitability of the operation for the building owner in the multi-objective optimization. The opposite approach of Sani Hassan \cite{SaniHassan2018} is to optimize the sizing and operation of an energy system on typical days and to study the impact on the voltage profile under various technological scenarios. The scenarios apply only to a single multi-family building while the load of the other bus remains identical. The work of Wang et al. \cite{Wang2018} proposes to integrate the revenue from frequency control ancillary services and reliability services into the objective function. The model quantifies the value of such services.\\
The previously mentioned literature does not consider the use of advanced tariff structures nor their impact on the system configurations. Schreiber et al. \cite{Schreiber2015} propose electricity tariffs and evaluate their impact on the grid requirements for two different buildings under two technological scenarios. Deetjen et al. \cite{Deetjen2018} propose an integrated convex formulation for equipment sizing and optimal control for a central utility plant in order to better integrate the surrounding rooftop PV. The authors consider time-of-use electricity pricing to be the best solution for providing flexibility services, absorbing the excess PV generation. More tariff scenarios are considered in the study of Ren \cite{Ren2016}, which specially focuses on the profitability of PV and battery system for owners. The authors describe how capacity tariffs can lower the revenue of the owners because PV generation cannot efficiently reduce peak demand. 
The previously mentioned studies are missing a few considerations, such as accounting for more than a single building,  allowing the PV and battery size to adapt to the scenarios, or integrating their assessment of building performance and network impact. There is thus a lack of literature considering the synergies between mitigating the network impact of PV systems and enabling profitable investment in PV-battery systems.

In this work, we investigate the effect of five different tariff scenarios, described in Table~\ref{tab:tarscen}, on network operation when optimizing both the design and operation of all buildings connected to that network, using a methodology developed in \cite{DACHpaper}. The scenarios consider pure volumetric electricity tariffs, a mix of volumetric and capacity-based tariffs, or a block rate tariff.  The optimization is run for a set of buildings in a sub-network of Rolle (Switzerland). The buildings' characteristics are known from a geographical information system. The resulting loads and generations at each injection point allow for solving the power flow equations over a full year to extract the distribution of the voltage level and line loading. Finally, these distributions are compared with a reference case to assess the effect of the tested tariffs.

\section{Methodology}
This section briefly summarizes the equations of the optimization problem, defines the performance metrics for the design and operation of all buildings, and presents the performance metrics from the grid perspective. 

\subsection{PV-battery optimal sizing and operation}
The PV and battery sizing and operation are optimized for each building to minimize the total cost of ownership given a set of modeling constraints \cite{DACHpaper}. The objective function includes both the investment and operational cost, as described in  \eqref{eq:objfcn}. By definition, the optimization problem relies on the assumption that an exact forecast of both the PV generation and the electrical load is provided. The impact of forecast errors and energy managers' performance is outside the scope of this study. In the following, $P$ denotes a power, $C$ a cost and $L$ a duration. The complete definitions of all variables are available in \cite{DACHpaper}.

\begin{align}
\label{eq:objfcn}
    \text{obj} &= \frac{r \cdot (1 + r) ^L}{ (1 + r) ^ L - 1} \cdot \left(\textsc{cx}_\text{pv} + \frac{L}{L^\textsc{bat}}\cdot \textsc{cx}_\text{bat}\right) + \textsc{opex}\\
\label{eq:cxPV}
\textsc{cx}_\text{pv} &=  \sum_{i=1}^N n^\textsc{mod}_i \cdot P^\textsc{mod}_{\text{nom},i} \cdot C^\textsc{mod}_i + b^\textsc{pv} \cdot \textsc{Cf}^\textsc{pv} \\
\label{eq:CXbat}
\textsc{cx}_\text{bat} &=  E^\textsc{bat}_\textsc{cap}\cdot C^\textsc{bat} + b^\textsc{bat}\cdot C_F^\textsc{bat}
\end{align}

\noindent where the decision variables are the number of PV modules installed $n^\textsc{mod}_i$ for each configuration, and the battery size $E^\textsc{bat}_\textsc{cap}$. The boolean variables $b^\textsc{pv}, b^\textsc{bat}$ are constrained to switch from 0 to 1 if the corresponding capacity is greater than 0.

\begin{figure}[!ht]
\centering
\includegraphics[width=\figWidth]{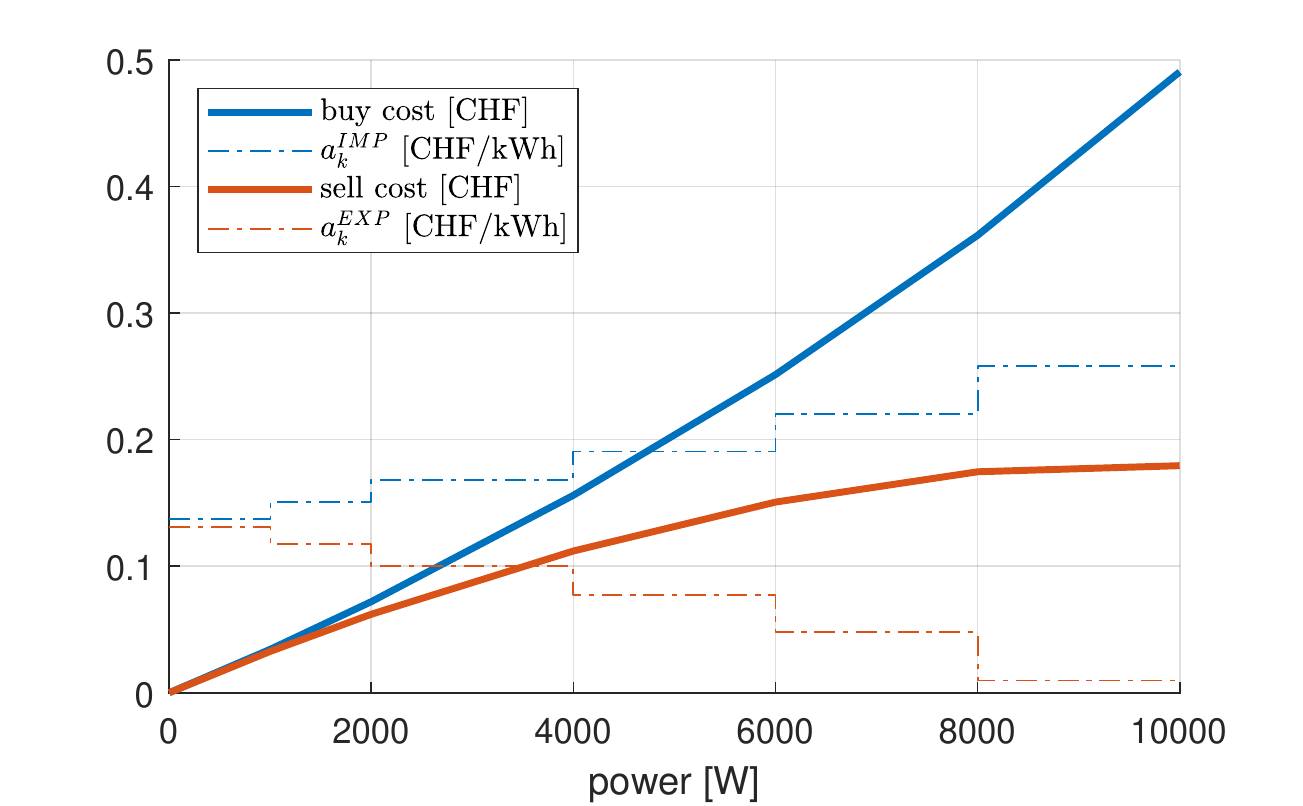}
\caption{Block rate tariff.}
\label{fig:blockrate}
\end{figure}

The operation decision variables are the battery charging and discharging power ($P^\textsc{cha,dis}_{t}$) and the PV curtailment ($P^\textsc{cur}_{t}$). Through the conservation of energy \eqref{eq:Econserv}, these decision variables determine the power withdrawn from - injected to -  the grid ($P^\textsc{imp}_{t}, P^\textsc{exp}_{t}$). Then, the operating cost $\textsc{opex}$ can be evaluated, considering the sum of the grid exchange cost $ \textsc{ox}_\text{ge}^\text{st}$ according to the selected tariff structure (volumetric, capacity, or block rate). The operating cost also contains the PV maintenance cost $\textsc{ox}_\text{pm}$ as defined in \eqref{eq:opex}.  

\begin{equation}
\label{eq:Econserv}
\begin{split}
P^\textsc{imp}_{t} - P^\textsc{exp}_{t} -  P^\textsc{cha}_{t} +  P^\textsc{dis}_{t}  - P^\textsc{cur}_{t} +P^\textsc{pv}_t \\
= P^\textsc{load}_t \quad \forall t\in T
\end{split}
\end{equation}
\noindent where  $P^\textsc{pv}_t = \sum_{i=1}^N P^\textsc{mod}_{t,i}\cdot n^\textsc{mod}_i$ in which  $P^\textsc{mod}_{t,i}$ is the power generated by a single module of the $i^{th}$ configuration at time $t$.

\begin{subequations}
    \begin{align}
    \label{eq:opex}
        & \text{Op. cost} &\textsc{opex} &= \sum_{st}^{\left[vol,pow,block\right]} \textsc{ox}_\text{ge}^\text{st}  +\textsc{ox}_\text{pm}\\
    \label{eq:pvm}
        & \text{PV maint.} & \textsc{ox}_\text{pm} &= \gamma^{PV} \cdot \textsc{cx}_\text{pv} \\
    \label{eq:ge_vol}
        & \text{Volumetric } & \textsc{ox}_\text{ge}^\text{vol} &= \sum_{t=1}^T \left[ P^\textsc{imp}_t \cdot  t^\textsc{imp}_t  - P^\textsc{exp}_t \cdot  t^\textsc{exp}_t\right ]\cdot \textsc{ts}_t\\
    \label{eq:ge_cap}
        & \text{Capacity } & \textsc{ox}_\text{ge}^\text{pow} &= \sum_{m=1}^M  P^\textsc{max}_{m} \cdot  t^\textsc{max}\\
    \label{eq:ge_stp}
        & \text{Block rate } & \textsc{ox}_\text{ge}^\text{block} & = \sum_{t=1}^T \max_{k=1\dots K} \left( P^\textsc{imp}_t \cdot a_k^\textsc{imp}\cdot \textsc{ts}_t  +b_k^\textsc{imp}\right) \nonumber\\
        & & & \quad - \sum_{t=1}^T \min_{k=1\dots K} \left( P^\textsc{exp}_t \cdot a_k^\textsc{exp}\cdot \textsc{ts}_t +b_k^\textsc{exp}\right)
    \end{align}
\end{subequations}
\noindent where the maximum power for month $m$, $P^\textsc{max}_{m}$, is calculated by requiring that both the import and the export power be smaller than this variable. The values of the import and export tariff $t^\textsc{imp}_t,t^\textsc{exp}_t$, the value of the capacity tariff $t^\textsc{max}$ and the values of the block rate tariff coefficients $a_k^\textsc{imp},a_k^\textsc{exp}$ are given in  Table~\ref{tab:tarscen}. The values of the coefficients $b_k^\textsc{imp}$ and $b_k^\textsc{exp}$ are calculated to ensure the continuity of the cost of buying or selling energy to the grid (see Fig.~\ref{fig:blockrate}). All other variables from \eqref{eq:objfcn} to \eqref{eq:opex} which are known parameters of the optimization problem are defined in  Table~\ref{tab:param}. 

\subsection{Performance metrics}
The performance metrics aim to assess the system's reaction, in terms of equipment size and operation, and the network's reaction, in terms of voltage profile and line loading, when changing the electricity pricing structure.  From a design perspective, the $\text{PV host}$ \eqref{eq:pvhost} is the ratio between the installed PV capacity and the maximum PV potential capacity of the building. The $\text{PV penetration}$ \eqref{eq:pvpenetr} compares the energy generated by the PV arrays with the annual consumption. The battery autonomy ratio, $\text{BAT auto}$ \eqref{eq:batauto}, corresponds to the ratio between the battery capacity and the mean daily consumption of the building. This metric can be understood as the fraction of a day that can be covered by the battery in the event of a blackout.
From an operation perspective, the $\text{PV cur}$ \eqref{eq:PVcur} is the fraction of the energy that is curtailed from the PV generation. The self-sufficiency $\text{SS}$ \eqref{eq:ss} is the fraction of the energy consumption that is self-covered by the PV-battery system. The definition of \eqref{eq:ss} is derived from \cite{Luthander2015}. 
To assess how the buildings interact with the grid, we defined in \cite{DACHpaper}  a grid usage ratio,  \text{GU IMP,EXP} \eqref{eq:GU}, as the ratio between the maximum withdrawn/injected power and the maximum load. Finally, from an economic perspective, the payback period, $\textsc{dpp}$ \eqref{eq:DPP}, (time to recover the investment) is of crucial interest to evaluate the profitability of the proposed economic framework. The levelized cost of energy ($\textsc{LCOE}$ \eqref{eq:LCOE}) also helps to assess whether the chosen scenario induces an increase or a decrease in the electricity price. 

\begin{subequations}
\label{eq:Metrics}
\begin{align}
\label{eq:pvhost}
     \text{PV host}&=\textsc{pv}^\textsc{cap}/\textsc{pv}^\textsc{cap}_\text{max}\\
\label{eq:pvpenetr}
     \text{PV penetration}&=\sum_t P_t^\textsc{pv}/\sum_{t} P_t^\textsc{load}\\
\label{eq:batauto}
      \text{BAT auto}&=\frac{E^\textsc{bat}_\textsc{cap}}{\text{mean daily energy}}\\
\label{eq:PVcur}
    \text{PV cur}&= \sum_t P^\textsc{cur}_{t}/\sum_t P^\textsc{pv}_t \\
\label{eq:GU}
    \text{GU IMP,EXP} &= \max(P^\textsc{imp,exp}_t)/\max(P^\textsc{load}_t)\\
\label{eq:NPV}
    \textsc{NPV} &= \sum_{t=1}^L \textsc{cf}_t/\left(1+r\right)^t\\
\label{eq:DPP}
    \textsc{DPP} &= T \mid \frac{\sum_{t=1}^T \textsc{cf}_t -\textsc{opex}_t^0}{\left(1+r\right)^t}=0 \\
\label{eq:LCOE}
    \textsc{LCOE} &= \frac{\textsc{npv}}{L \cdot \sum_t P^\textsc{load}_t\cdot TS_t}
\end{align}
\begin{equation}
\label{eq:ss}
     \textsc{SS} = \frac{\sum_{t} \min(P_t^\textsc{load},P_t^\textsc{exp}+P_t^\textsc{load}-P_t^\textsc{imp})}{\sum_{t} P_t^\textsc{load}}
\end{equation}
\end{subequations}

\noindent where $\textsc{cf}_t$ is the net cash flow (investment + maintenance cost + operational cost, including battery replacement) at time t and $\textsc{opex}_t^0$ is the original operating cost without the investment in the PV and battery. 

The resolution of the power-flow equations allows for extracting the voltage (in per unit, p.u) at every node of the network and the current flowing through every line. Given the lines' properties, in particular, the maximum allowable current, a representative metric for grid congestion is the line loading level. We consider separately the situation when the bus voltage at an injection point is above 1 p.u and when it is below 1 p.u, and use the 95th percentile of the bus voltage deviation. This allows us to distinguish when there is a local excess of energy from when there is a local deficit of energy. Finally, one of the key issues for the high penetration of distributed stochastic generators is the reverse-power flow occurring at the link between the low-voltage side and the upper level. 
For this reason, the load duration curve enables us to assess the requirement in terms of power that has to flowed into and out of the low-voltage grid.

\section{Case study}
\begin{figure}[!ht]
\centering
\includegraphics[width=.8\linewidth]{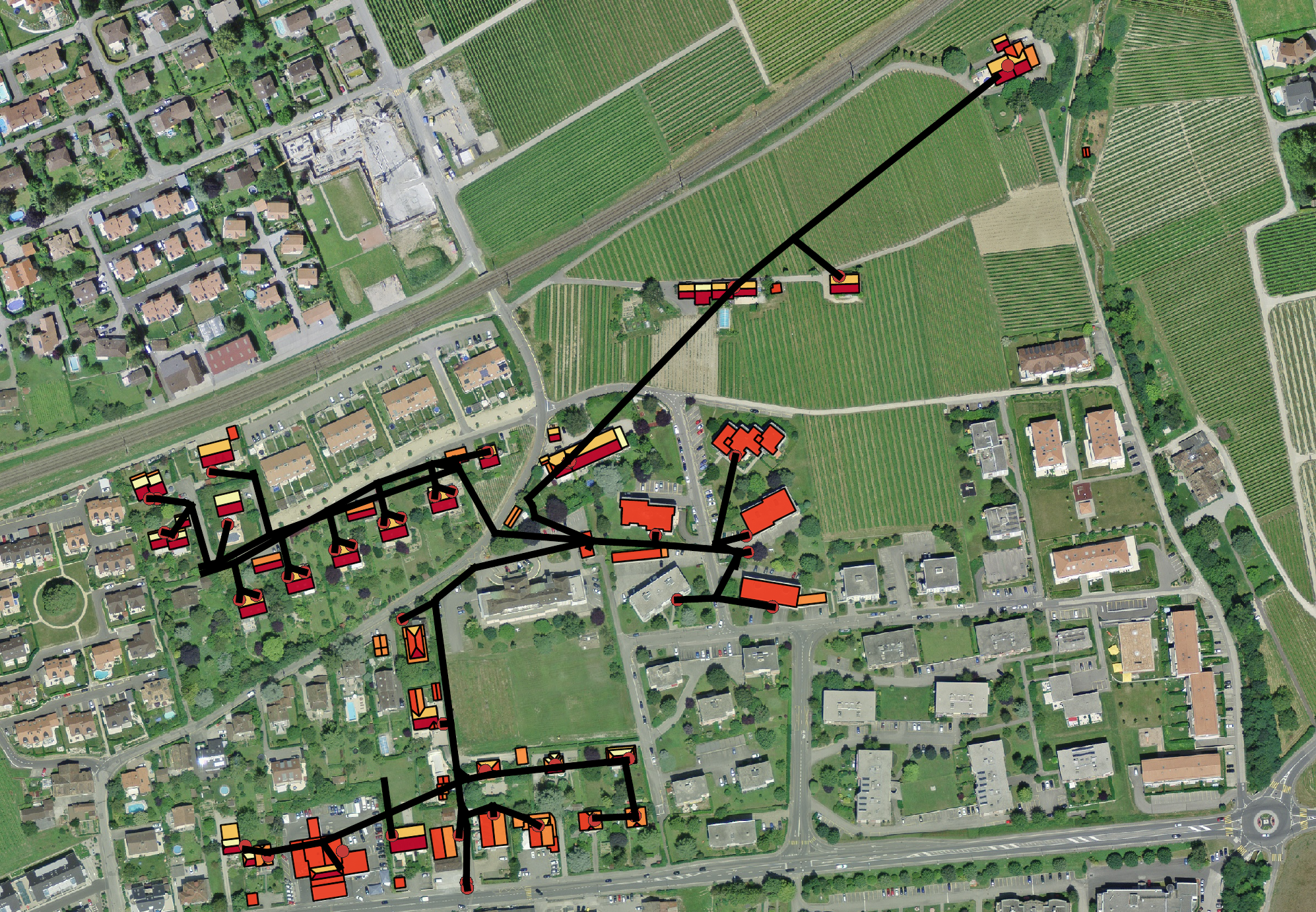}
\caption{A map of the considered low-voltage grid (black lines). The roofs are colored according to their mean annual solar irradiance.}
\label{fig:net}
\end{figure}

\begin{figure}[!ht]
\centering
\includegraphics[width=\figWidth]{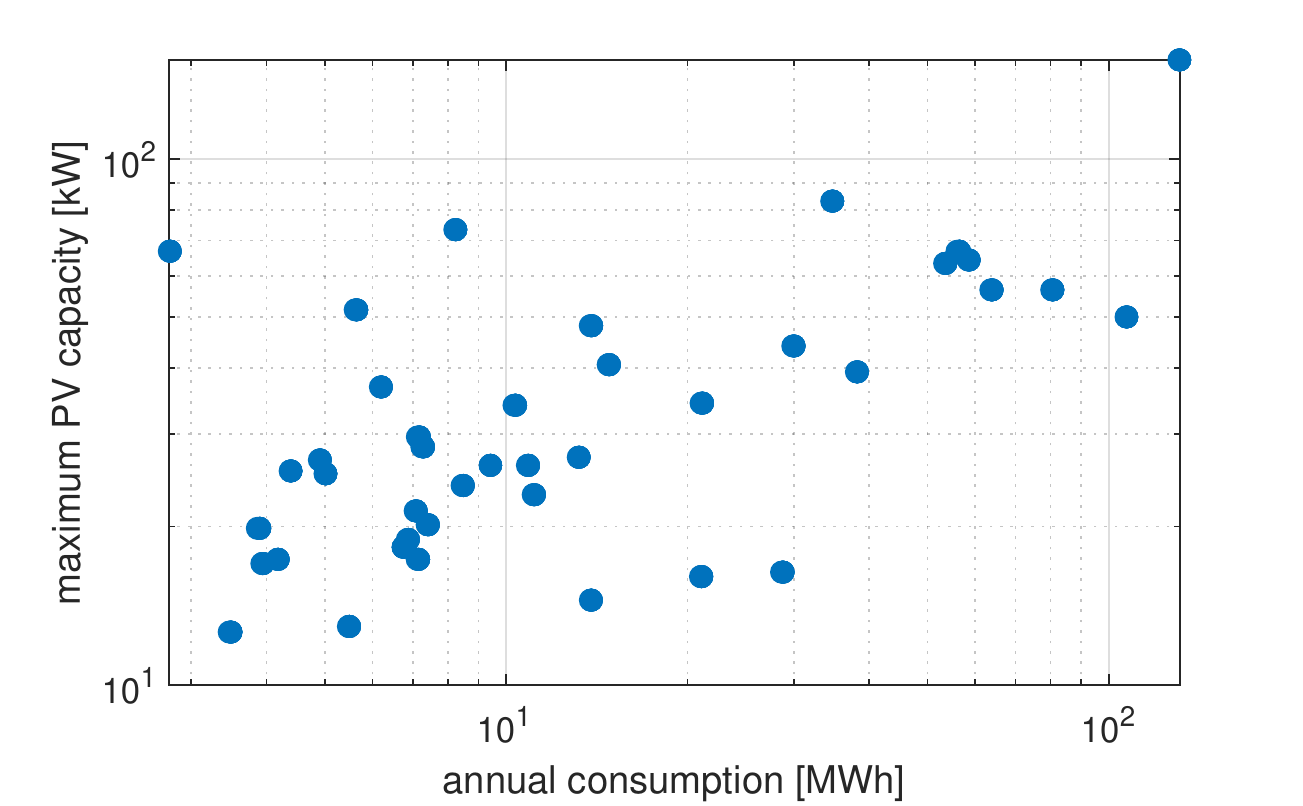}
\caption{Potential PV capacity and annual consumption plot.}
\label{fig:ConsPVcapMal}
\end{figure}

The methodology is applied to a case study in Rolle (Switzerland) where a low-voltage distribution grid has been modeled (Fig.~\ref{fig:net}). The grid hosts 41 buildings. The properties of each building are known thanks to publicly available geographical information system\footnote{\url{https://www.asitvd.ch/}}. The annual consumption for each meter in the grid was provided, allowing us to match the meters with real smart-meter measurements and allocate the smart-meter time-series to each meter. The roof's surface area, azimuth, and tilt are known for each building. With this, we are able to calculate, for each roof and then for each building, the maximum potential PV capacity as shown in Fig.~\ref{fig:ConsPVcapMal}. The latter figure shows the maximum PV capacity as a function of the building's annual consumption. 

Five tariff scenarios are defined in Table~\ref{tab:tarscen}. The first three scenarios consider volumetric tariffs in which the costs/revenues are proportional to the exchanged energy according to a given tariff. The first scenario is a reference tariff that is constant throughout the day. The second scenario, called "solar tariff", promotes consumption when solar irradiance is higher by setting a low energy rate during 11h-15h. The third scenario mirrors the spot market price (actually the intraday continous from the EPEX market data\footnote{EPEX price for 2018 \url{https://www.epexspot.com/en/market-data/intradaycontinuous/intraday-table/-/CH}}). The fourth scenario is a mix of a volumetric and capacity-based tariff, in which the cost is proportional to the monthly maximum power exchanged with the grid and the energy consumed from the grid, while the revenue is proportional to the energy injected into the grid. The fifth scenario considers a block rate tariff, in which the cost/revenue is proportional to the energy exchanged with the grid, but the tariff depends on the power level at which the energy is exchanged. All these tariffs have been calibrated such that the total cost/revenue for the system operator remains approximately the same. 

\begin{table}[!ht]
\renewcommand{\arraystretch}{.9}
\centering
\caption{Tariff scenarios}
\label{tab:tarscen}
\begin{tabular}{@{}lrll@{}}
\toprule
\textbf{Scenario}                       & \textbf{Description} & \multicolumn{2}{l}{\textbf{Tariff} (cts/kWh unless specified)}\\
\midrule
\multirow{2}{*}{\textbf{Reference}}    & $t^\textsc{imp}_t$:              & 21.02          &                    \\
                                       & $t^\textsc{exp}_t$:              & 8.16           &                    \\
                                       \midrule
\multirow{4}{*}{\textbf{Solar}} & $t^\textsc{imp}_{t\in 11h:15h}$:       & 14.68          &                    \\
                                       & $t^\textsc{exp}_{t\in 11h:15h}$:       & 7.07           &                    \\
                                       & $t^\textsc{imp}_{t\notin 11h:15h}$:    & 23.17          &                    \\
                                       & $t^\textsc{exp}_{t\notin 11h:15h}$:    & 11.12          &                    \\
                                       \midrule
\multirow{2}{*}{\textbf{Spot market}}   & $t^\textsc{imp}_t$: & \multicolumn{2}{l}{EPEX*3.9468}                      \\
                                       & $t^\textsc{exp}_t$:              & \multicolumn{2}{l}{EPEX*1.604}                      \\
                                       \midrule
\multirow{3}{*}{\textbf{Capacity}}     & $t^\textsc{imp}_t$:            & 15.91          &                    \\
                                       & $t^\textsc{exp}_t$:            & 12.09          &                    \\
                                       & $t^\textsc{max}$:              & \multicolumn{2}{l}{5.02 CHF/kW/month}                    \\
                                       \midrule
\multirow{7}{*}{\textbf{Block rate}}   & Power (kW)           & $a_k^\textsc{imp}$    & $a_k^\textsc{exp}$     \\
                                       & 0 to 1               & 13.72                 & 13.07              \\
                                       & 1 to 2               & 15.06                 & 11.73              \\
                                       & 2 to 4               & 16.80                 & 9.99               \\
                                       & 4 to 6               & 19.07                 & 7.73              \\
                                       & 6 to 8               & 22.01                 & 4.79               \\
                                       & 8 to 10              & 25.83                 & 0.96              \\
                                \bottomrule
\end{tabular}
\end{table}

The technology price levels have been set according to the reference year 2025. The price projections have been extracted from the IRENA report \cite{IRENA2017} and calibrated for the Swiss price levels provided in a recent market study \cite{Planair2019}. All parameters of the problem are provided in  Table~\ref{tab:param}.

\begin{table}[!ht]
\renewcommand{\arraystretch}{.9}
\centering

\caption{Parameters}
\begin{threeparttable}
\begin{tabular}{@{}lllp{3.5cm}@{}}
\toprule
 & \textbf{Parameter}  & \textbf{Value} & \textbf{Description}  \\
\midrule
\parbox[t]{2mm}{\multirow{3}{*}{\rotatebox[origin=c]{90}{\textsc{Time}}}}
&$T$ & 35040 & number of time steps \\ 
&$M$ & 12 & number of months\\ 
&$\textsc{ts}$  & 900\,s & time steps \\ 
\midrule
\parbox[t]{2mm}{\multirow{6}{*}{\rotatebox[origin=c]{90}{\textsc{PV}}}}
&$N$  & \tnote{1} & number of PV configurations \\ 
&$\textsc{Cf}^\textsc{pv}$  & 10049\,CHF & PV fixed cost\\  
&$C^\textsc{mod}$  & 1.05\,CHF/W  & PV configurations, specific costs \tnote{2} \\ 
&$P^\textsc{mod}_\text{nom}$ & 315\,W  & PV unit nominal power \tnote{2}\\
&$P^\textsc{mod}_{t,i}$  &  \tnote{3} & PV configuration unit generation\\
&$\gamma^{PV}$ & 0.5\% & annual maintenance specific cost \\
\midrule
\parbox[t]{2mm}{\multirow{2}{*}{\rotatebox[origin=c]{90}{\textsc{Batt}}}}
&$C^\textsc{bat}$  & 229\,CHF/kWh & battery specific cost \\ 
&$C_F^\textsc{bat}$ & 0\,CHF & battery fixed cost \\  
\midrule
\parbox[t]{2mm}{\multirow{3}{*}{\rotatebox[origin=c]{90}{\textsc{Other}}}} 
&$L$ &  25 years & system lifetime \\
& $L^\textsc{bat}$  & 9 years & expected battery lifetime \\
& $r$ &  3\% & discount rate \\ 
\bottomrule
\end{tabular}
\begin{tablenotes}
    \item[1] data from the geographical information system
    \item[2] unique value for all configuration and roofs
    \item[3] PV profiles are simulated using the PVLIB toolbox \cite{stein_pvlib:_2016}
\end{tablenotes}
\end{threeparttable}
\label{tab:param}
\end{table}



\section{Results and discussion}
The optimizations of the 41 buildings were performed on a Intel(R) Xeon(R) CPU E5-2630 v3 @ 2.40GHz processor with 8 Cores and 32GB of RAM using  \textsc{gurobi}\cite{gurobi} to solve the mixed-integer-linear problem. The power flows were then solved for each time step with a resolution of 15\,min using \textsc{Pandapower} \cite{pandapower.2018}. 

\subsection{Design and operation of the PV-battery energy systems}
\begin{figure}[!ht]
\centering
\includegraphics[width=\figWidth]{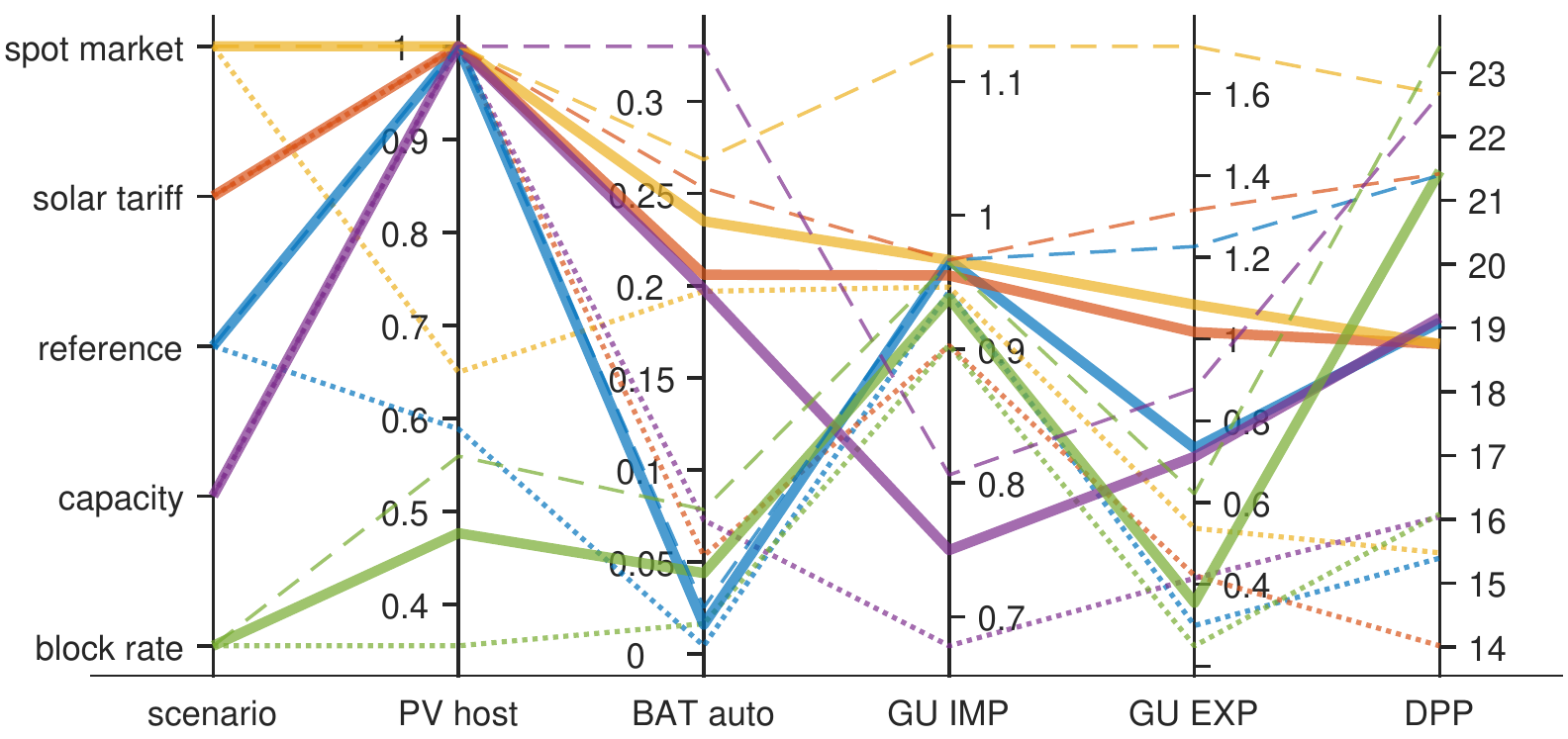}
\caption{PV hosting ratio, battery autonomy, grid usage ratio (import/export) and discounted pay back period for all scenarios. Metrics are defined in \eqref{eq:Metrics}. Solid lines are the median, dashed lines are the 75th percentile, and dotted lines are the 25th percentile.}
\label{fig:parallel}
\end{figure}

The resulting designs are pictured in Fig.~\ref{fig:parallel}. In all scenarios except the block rate tariff, almost all the roofs are covered with PV leading to a PV host value close to one. The block rate scenario, however, limits the penetration of PV. Regarding battery size, investment in such technology is driven by economic opportunities, namely by variations in the electricity price (solar and spot market tariff scenarios) or by a strong incentive to limit the exchanged power (capacity tariff scenario). Although this last aspect is also present in the block rate scenario, the incentive is, thus, lower, leading to lower relative battery size. In terms of grid usage, only the capacity tariff and block rate tariff scenarios provide a clear incentive to reduce the maximum power exchanged with the grid. The spot market and solar tariff scenarios, due to the volatility of electricity prices, tend to increase the power injected or withdrawn from the grid. On the economic side, the discounted payback periods are similar, ranging from 14 to 23 years for all scenarios. The median is, however, higher for the block rate tariff, reaching 21 years against 19 years for the other scenarios. 

\begin{figure}[!ht]
\centering
\includegraphics[width=\figWidth]{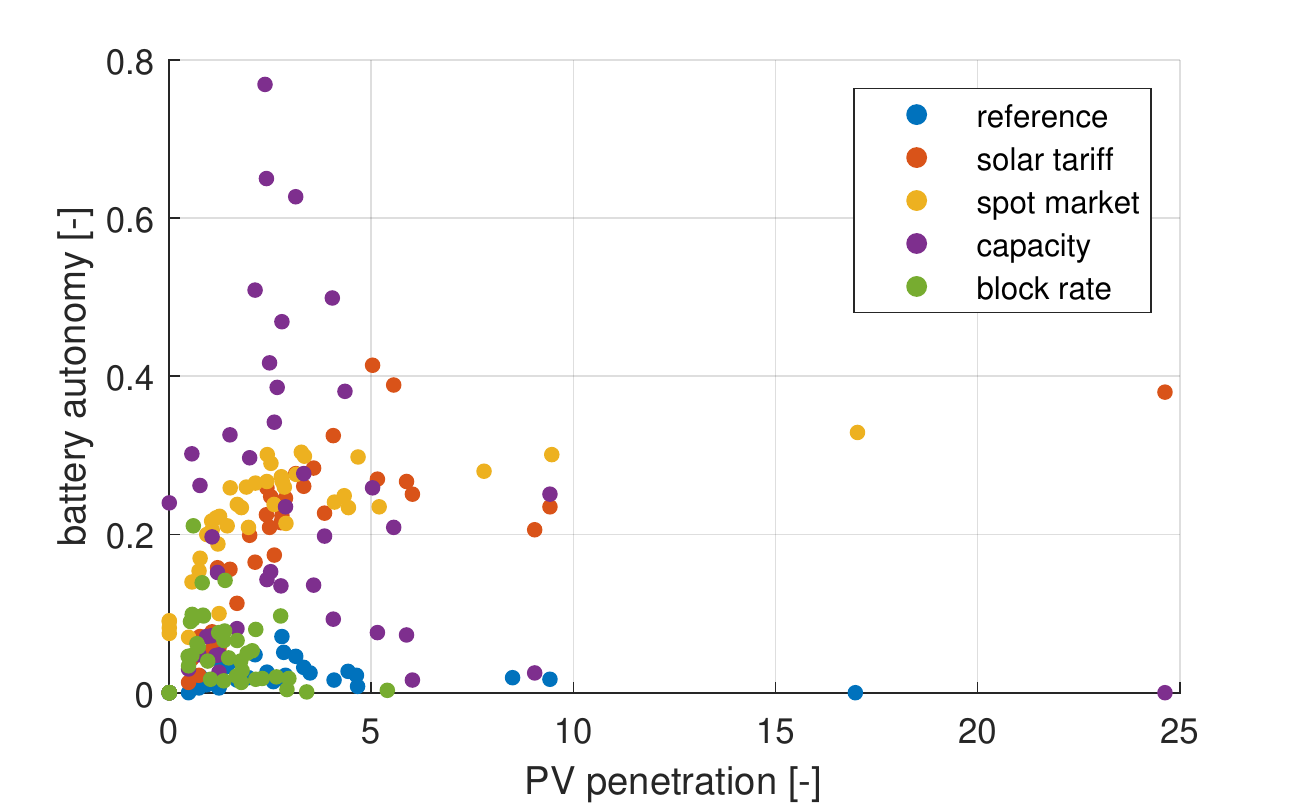}
\caption{Battery autonomy versus PV penetration.}
\label{fig:BatAutoVsPVpenetr}
\end{figure}

\begin{figure}[!ht]
\centering
\includegraphics[width=\figWidth]{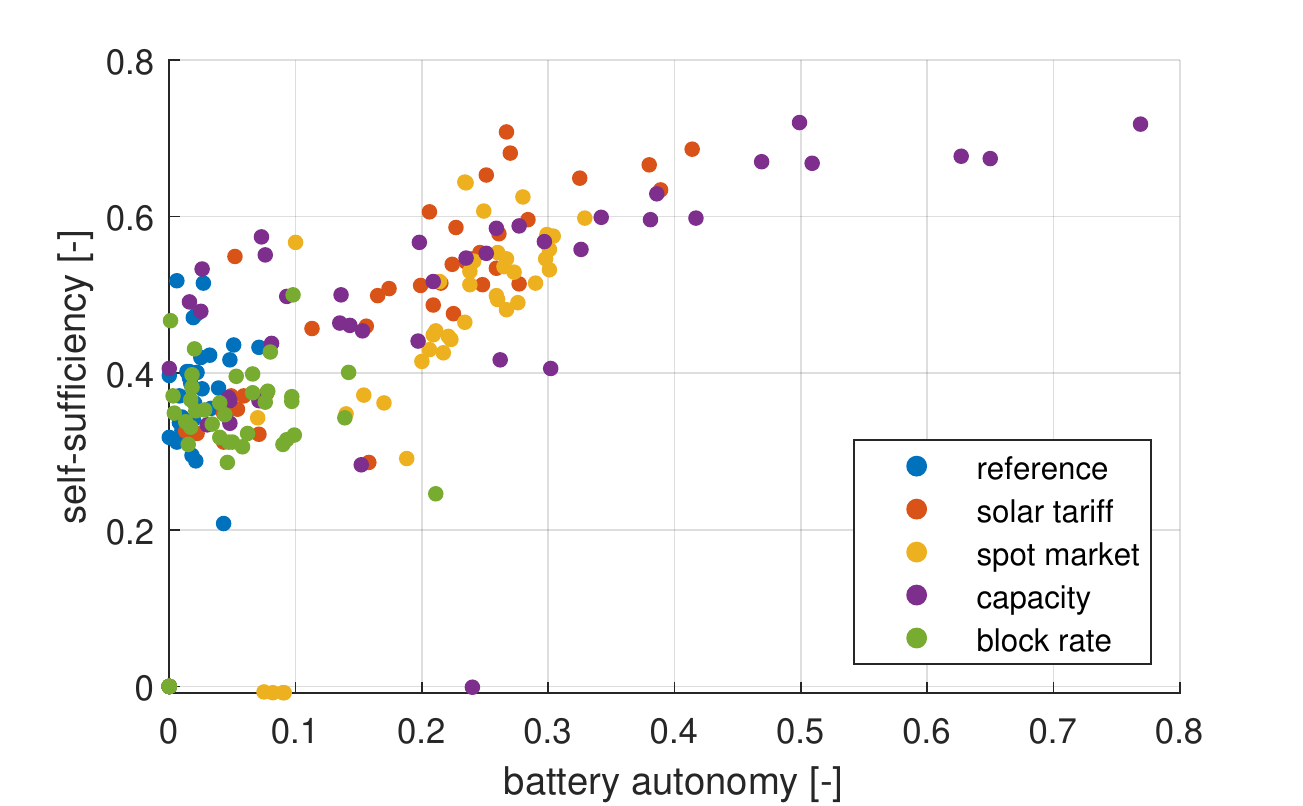}
\caption{Self-sufficiency level against the battery autonomy ratio.}
\label{fig:SSvsBatAuto}
\end{figure}

\begin{figure}[!ht]
\centering
\includegraphics[width=\figWidth]{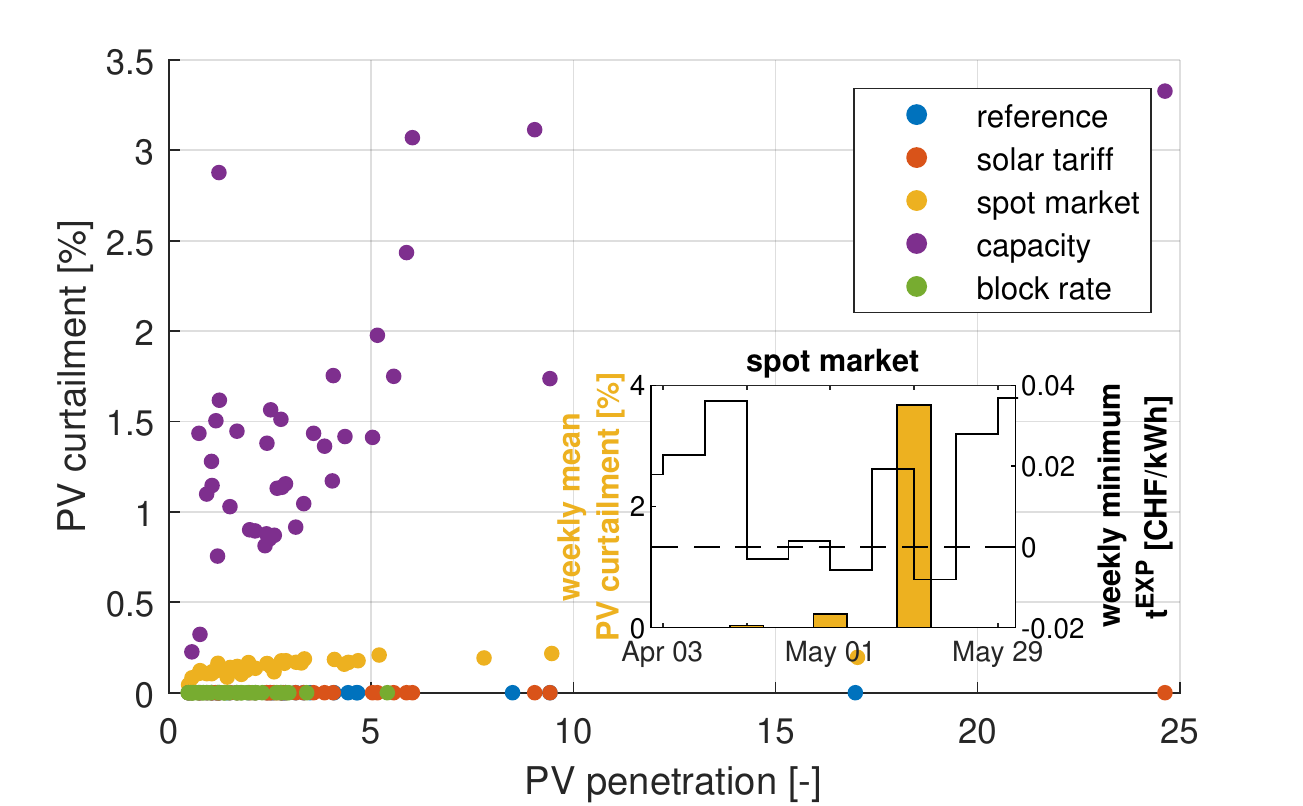}
\caption{Ratio of energy curtailed and PV penetration. In the inset, the bars are the weekly ratio of energy curtailed (left axis) and lines the weekly minimum $t^\textsc{exp}$ (right axis).}
\label{fig:PVcurVsPVcap}
\end{figure}

The relative size of the battery does not scale linearly with PV penetration, as depicted in Fig.~\ref{fig:BatAutoVsPVpenetr}, except for the dynamic volumetric tariffs (spot market and solar). For the capacity and block rate tariffs, low PV penetration, underlying a high consumption level compared to the PV production, tends to increase the battery autonomy ratio in order to limit the import power. Conversely, at high PV penetration, the battery autonomy tends to decrease for the capacity and block rate tariffs. For the first case, the role of the battery to cut injection peaks is replaced by the curtailment of the PV generation (Fig.~\ref{fig:PVcurVsPVcap}). As curtailing is free, there is no need to invest in batteries for this purpose. For the block rate scenario, high injection is not penalized; the marginal revenue is just decreased. Thus, it limits the profitability of having a high PV capacity compared to its consumption level, but does not require either curtailment of the PV energy or investing in storage technologies. The fraction of energy curtailed is zero for all scenarios except the capacity and spot market tariffs. For the latter, the small fraction of curtailed energy is due to negative spot market prices as shown in the inset of Fig.~\ref{fig:PVcurVsPVcap}. 
As a general trend, a larger battery size relative to the building consumption increases the self-sufficiency ratio as shown in Fig.~\ref{fig:SSvsBatAuto}. This trend is very pronounced for the spot market, solar and capacity tariffs, although a saturation appears for the latter.  

    

\begin{figure}[!ht]
\centering
\includegraphics[width=\figWidth]{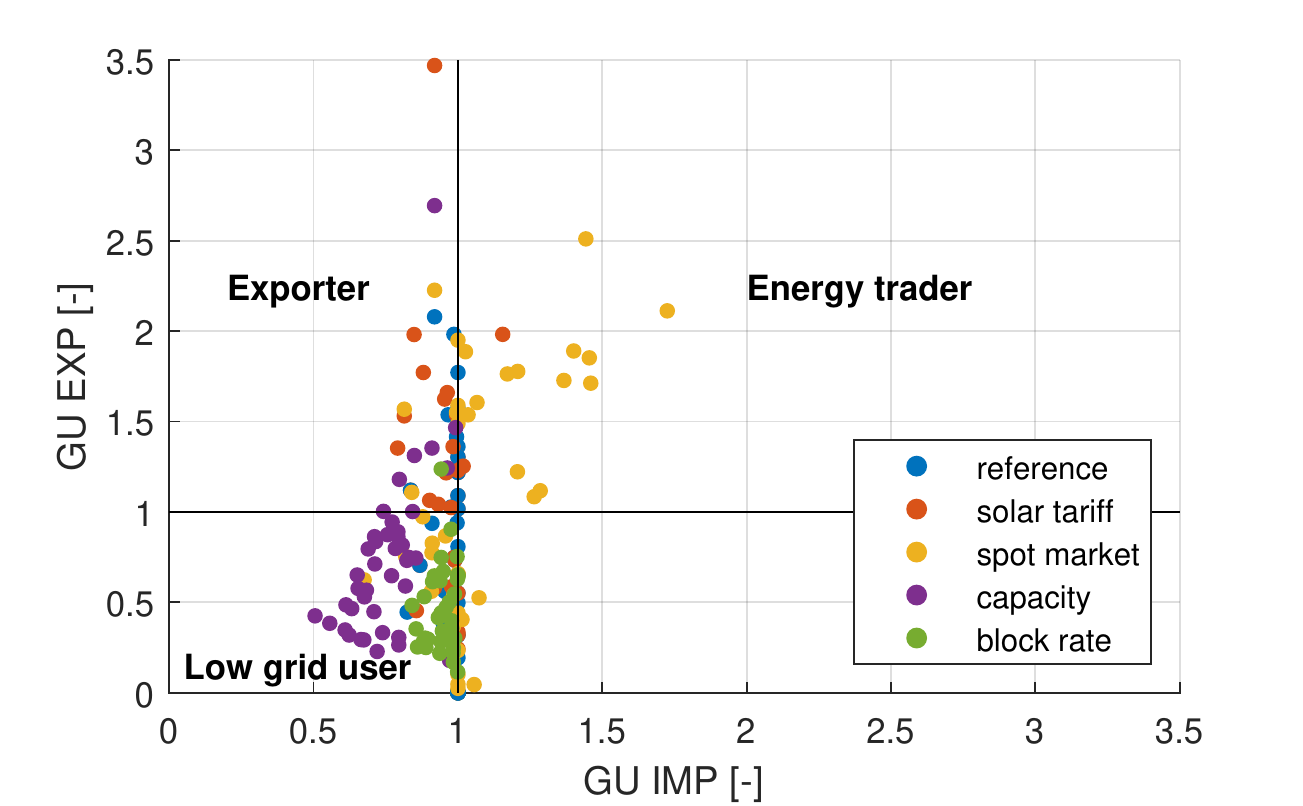}
\caption{Export power ratio vs import power ratio.}
\label{fig:PSexpPSimp}
\end{figure}

In summary, compared to the reference scenario, dynamic volumetric tariffs (solar and spot market) promote investments in storage technologies since they provide economic opportunities to generate profit for the building owners. The capacity tariff promotes investment in storage, but the main function this is to reduce consumption peaks by curtailment. The block rate tariff promotes smaller PV penetration (thus, PV capacity) and battery capacity but achieves a self-sufficiency level similar to the reference case. As pictured in Fig.~\ref{fig:parallel}, these considerations have an impact on grid usage behavior. In particular, the grid usage ratios are higher (regardless when importing or exporting) for both the spot market and solar tariff. It is especially pronounced for the spot market case. Conversely, capacity tariffs significantly drop the grid usage ratio for import, while the block rate tariff reduces both. Fig.~\ref{fig:PSexpPSimp} illustrates these observations. This figure allows us to distinguish between three types of grid user: the exporters, who have a grid usage ratio for export above 1 and even reduce their import grid usage by covering their own consumption; the energy traders, who buy or sell energy to maximize their profit and who tend to keep to their grid usage ratio for import and export above 1 (if the grid usage for export is below 1 it means that their own consumption dominates their generation capabilities); and the low grid users, who reduce both grid usage ratios, thus interacting less and less with the network. Almost all buildings fall in this category for the block rate scenario. The following will show how these local design and operation adaptations affect the network operational metrics.

\subsection{Low-voltage grid impact}
In order to have a complete overview of the grid reaction to the different tariff scenarios (in which PV installations and batteries are always present), two additional scenarios are added. The first considers only the original load without any investment in PV or batteries; the second considers that all roofs are covered with PV (regardless of the profitability of such a decision) but with no investment in batteries. These two scenarios set the upper and lower bounds to the grid impact metrics. 

\begin{figure}[!ht]
\centering
\includegraphics[width=\figWidth]{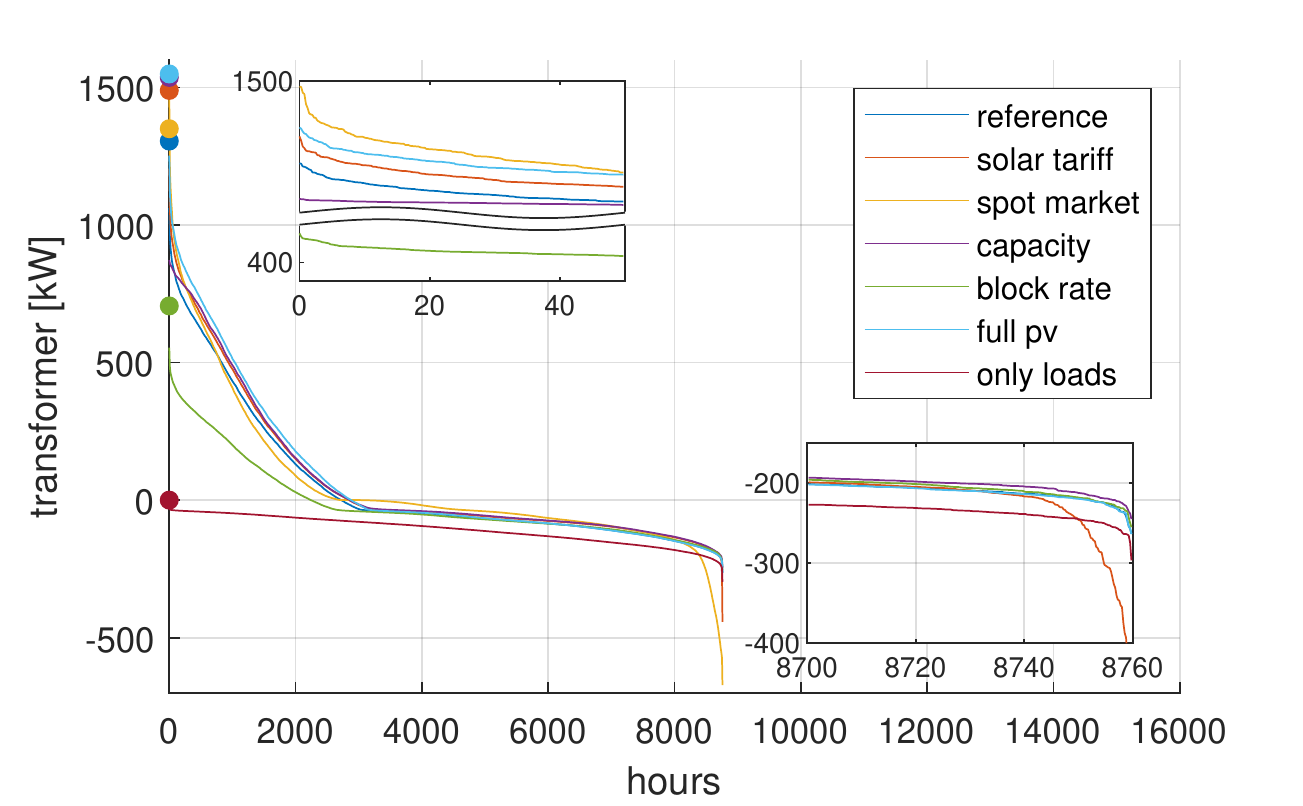}
\caption{Load duration curve at the transformer. Dots on the vertical axis indicate the total installed PV capacity per scenario. The nominal transformer capacity is 400 kW. Negative values indicate power flow from the high-voltage toward the low-voltage side.}
\label{fig:loadDurCurve}
\end{figure}

\begin{figure}[!ht]
\centering
\includegraphics[width=\figWidth]{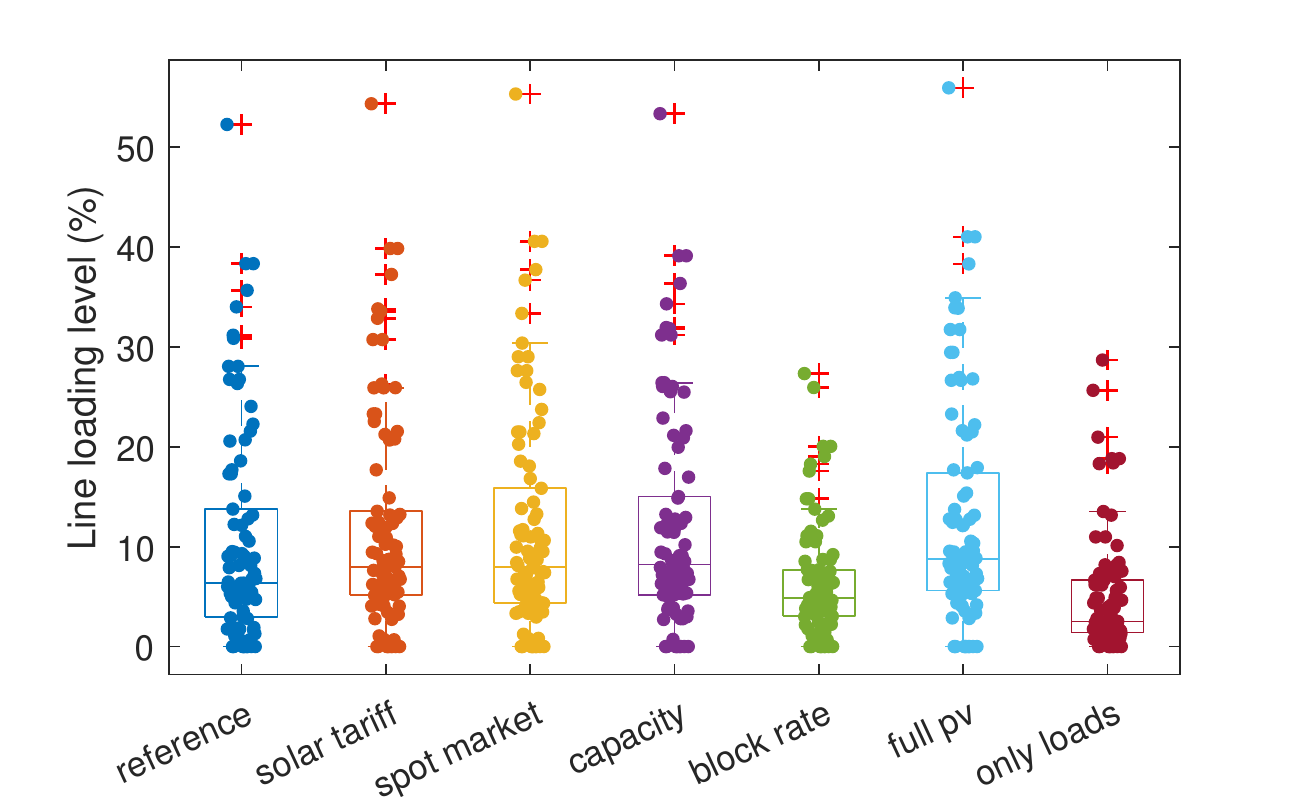}
\caption{95th percentile of the line loading level.}
\label{fig:lineLoading}
\end{figure}

The load duration curve in Fig.~\ref{fig:loadDurCurve} highlights the violation of the transformer power capacity for reverse power flow. All scenarios, except the load-only case, experience a maximum power flowing from the low-voltage side to the high-voltage side above 400 kW.  The block rate, helped by a significantly lower installed PV capacity, has the lowest maximum reverse power,  but a significant number of hours are above 400 kW. The most-demanding scenario is the spot market scenario which has the highest power demand and the highest injection power. The solar tariff also shows a significant increase in power demand compared to the other scenarios. This has direct consequences for the level of loading of the lines (ratio between current and the maximum nominal current of the line). Fig.~\ref{fig:lineLoading} shows that line loading level is significantly higher for all scenarios including PV, with the most extreme values attained under the spot market and full PV scenarios. The block rate tariff helps to significantly reduce the loading level of the lines. In this case, even the most loaded lines are less congested than in the load-only scenario.

\begin{figure}[!ht]
\centering
\subfloat[When above 1 p.u\label{fig:posVdev}]{\includegraphics[width=\figWidth]{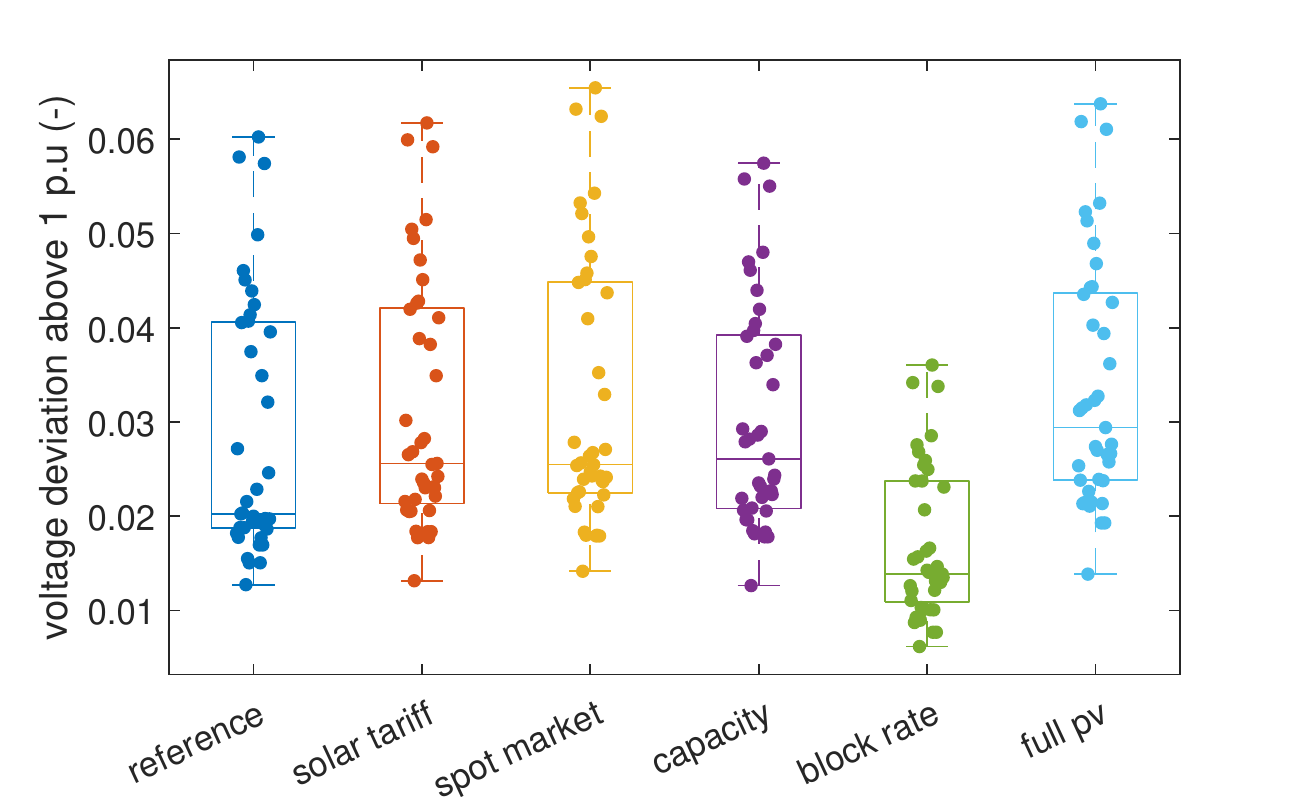}}\\[-.5ex]
\subfloat[When below 1 p.u\label{fig:negVdev}]{\includegraphics[width=\figWidth]{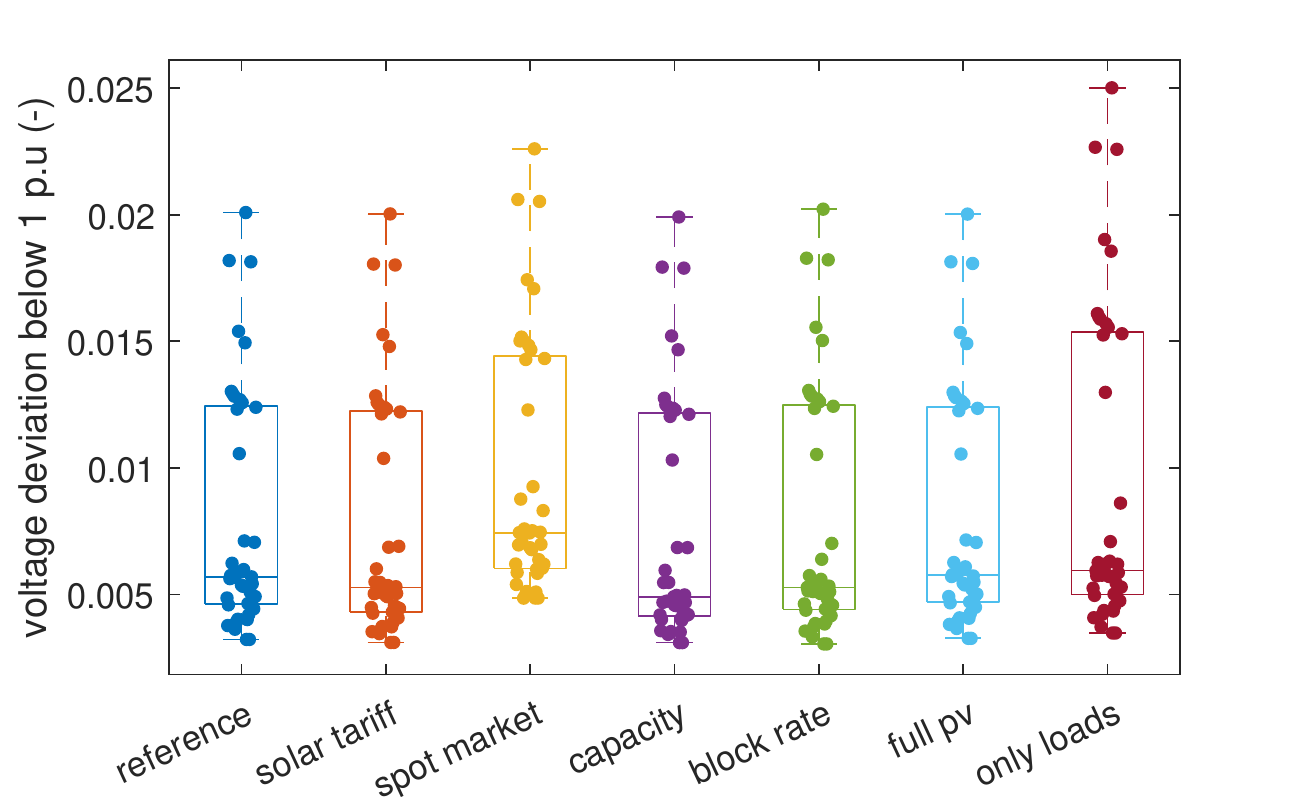}}
\caption{Voltage deviation distribution}
\end{figure}

One of the main concerns of grid operators regarding high PV penetration is to keep voltage levels to a value as close as possible to 1 p.u. As a matter of fact, scenarios, except the spot market scenario, fulfill the criteria EN50160, meaning the voltage levels fall within +-10\% of the nominal voltage for 95\% of each week.  When considering only the case when the voltage deviations exceed 1 p.u, Fig.~\ref{fig:posVdev} demonstrates the effectiveness of the capacity and block rate tariffs in terms of load management, as both lower the deviation of the most sensible bus compared to the reference and full PV case. Note that the load-only scenario is not displayed in this figure because the voltage levels never exceed 1 p.u. Alternatively, when the load level is below 1 p.u (Fig.~\ref{fig:negVdev}), only the spot market case significantly increases the voltage deviations. 

These observations highlight that advanced tariff structures can have two competing impacts. On the one hand, they help to mitigate the grid impact of distributed generation by promoting either small PV installations or moderate grid exchange. On the other hand, they can bring economic opportunities for significant investments in batteries and PV capacities but may increase the pressure on the grid. The most-concerning aspect is the transformer capacity to bring the excess power from the low-voltage side to the upper level. Over/under-voltage and overloading of the lines are in this case far less concerning. Although each scenario has been carefully calibrated to globally keep the same total cost (with respect to the reference case) for all buildings, it is worth investigating the levelized cost of energy (LCOE) served in order to make sure that the resulting systems do not actually suffer from a net increase in the energy price.   

\subsection{Economic aspects}

In the load-only scenario, the LCOE is 21 cts/kWh (corresponding to the import tariff of the reference scenario), while the LCOE can become higher in the full PV scenario, showing that an over-investment can occur and explaining why, in the reference case, some roofs are not covered with PV. Additionally, in the solar and spot market scenarios, only a small fraction of the buildings has an LCOE exceeding 21 cts/kWh, while the large majority would gain from switching to these tariff structures. For the capacity tariff scenario, a significant fraction of the buildings has an LCOE higher than the reference value, showing that, despite its positive impact on grid operation, such a tariff comes with a price for some building owners. The block rate scenario, though less prone for PV and battery investment, presents an LCOE lower than the reference values for all buildings. The design of such a tariff is a matter of compromise and there is an apparent lack of literature for designing a block rate tariff in the specific scope of promoting distributed renewable energy and mitigating grid impact. 

\begin{figure}[!ht]
\centering
\includegraphics[width=\figWidth]{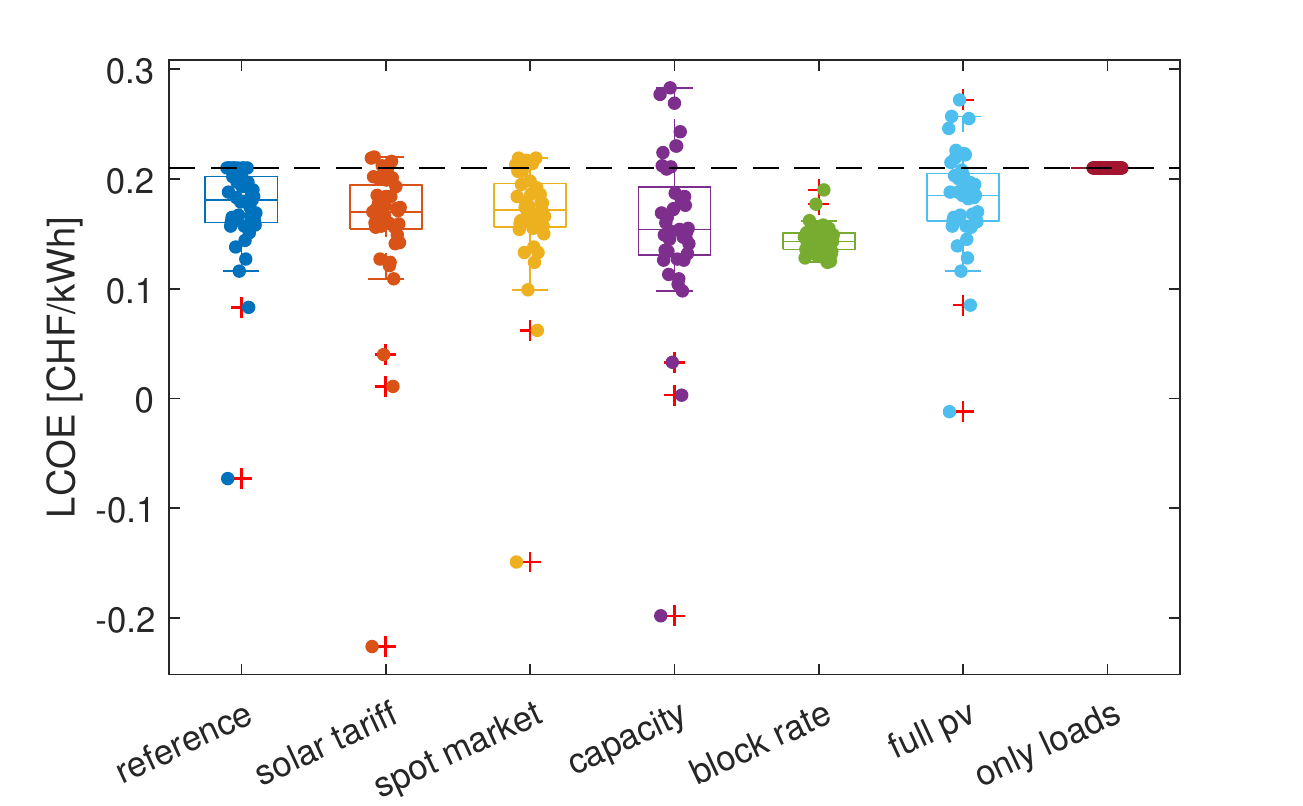}
\caption{Levelized cost of energy per scenario.}
\label{fig:LCOE}
\end{figure}



\section{Conclusion}
Our study performs the optimization of 41 buildings under five different tariff scenarios, including three volumetric tariffs, one combination of volumetric and capacity tariff and one block rate tariff. The resulting systems vary in term of installed PV capacity and battery capacity. The highest PV penetration is achieved with the capacity tariff while this scenario significantly reduces the voltage deviation and the line loading level. Volumetric tariffs, with high price volatility such as in the spot market scenario, lead to more investment and profitability of the batteries but also increase pressure on the network. Although the block rate scenario promotes smaller PV installation, it achieves the smallest median cost of providing energy for the end-users (from 18 cts/kWh in the reference case down to 14 cts/kWh). It reduces the 95th percentile of the positive voltage deviation of the bus with the larges deviation from 6\% to 3.6\% and reduces the maximum reverse power from 1060 kW to 550 kW, remaining above the 400 kW nominal power of the transformer. Further studies will elaborate on the design of block rate tariffs to mitigate network impact and incentivize high penetration of PV. The effect of a growing penetration of electric vehicles and heat pumps will also be considered. 




\bibliographystyle{IEEEtran}
\bibliography{biblio}

\begin{thebibliography}{10}
\providecommand{\url}[1]{#1}
\csname url@samestyle\endcsname
\providecommand{\newblock}{\relax}
\providecommand{\bibinfo}[2]{#2}
\providecommand{\BIBentrySTDinterwordspacing}{\spaceskip=0pt\relax}
\providecommand{\BIBentryALTinterwordstretchfactor}{4}
\providecommand{\BIBentryALTinterwordspacing}{\spaceskip=\fontdimen2\font plus
\BIBentryALTinterwordstretchfactor\fontdimen3\font minus
  \fontdimen4\font\relax}
\providecommand{\BIBforeignlanguage}[2]{{%
\expandafter\ifx\csname l@#1\endcsname\relax
\typeout{** WARNING: IEEEtran.bst: No hyphenation pattern has been}%
\typeout{** loaded for the language `#1'. Using the pattern for}%
\typeout{** the default language instead.}%
\else
\language=\csname l@#1\endcsname
\fi
#2}}
\providecommand{\BIBdecl}{\relax}
\BIBdecl

\bibitem{Tonkoski2012}
R.~Tonkoski, D.~Turcotte, and T.~H.~M. El-Fouly, ``{Impact of High PV
  Penetration on Voltage Profiles in Residential Neighborhoods},'' \emph{IEEE
  Transactions on Sustainable Energy}, vol.~3, no.~3, pp. 518--527, jul 2012.

\bibitem{Arboleya2017}
P.~Arboleya, M.~Huerta, B.~Mohamed, C.~Gonzalez-Moran, and X.~Dominguez,
  ``{Assessing the effect of nearly-zero energy buildings on distribution
  systems by means of quasi-static time series power flow simulations},'' in
  \emph{2017 IEEE Power {\&} Energy Society General Meeting}.\hskip 1em plus
  0.5em minus 0.4em\relax IEEE, jul 2017, pp. 1--5.

\bibitem{Hidalgo-Rodriguez2018}
D.~I. Hidalgo-Rodriguez and J.~Myrzik, ``{Optimal operation of interconnected
  home-microgrids with flexible thermal loads: A comparison of decentralized,
  centralized, and hierarchical-distributed model predictive control},''
  \emph{20th Power Systems Computation Conference, PSCC 2018}, 2018.

\bibitem{Hashemipour2018}
N.~Hashemipour, T.~Niknam, J.~Aghaei, H.~Farahmand, M.~Korpas, M.~Shafie-Khah,
  G.~J. Osorio, and J.~P. Catalao, ``{A linear multi-objective operation model
  for smart distribution systems coordinating tap-changers, photovoltaics and
  battery energy storage},'' \emph{20th Power Systems Computation Conference,
  PSCC 2018}, no. 309048, 2018.

\bibitem{SaniHassan2018}
A.~{Sani Hassan}, L.~Cipcigan, and N.~Jenkins, ``{Impact of optimised
  distributed energy resources on local grid constraints},'' \emph{Energy},
  vol. 142, pp. 878--895, jan 2018.

\bibitem{Wang2018}
H.~Wang, N.~Good, E.~A. Cesena, and P.~Mancarella, ``{Co-optimization of a
  multi-energy microgrid considering multiple services},'' \emph{20th Power
  Systems Computation Conference, PSCC 2018}, 2018.

\bibitem{Schreiber2015}
M.~Schreiber, M.~E. Wainstein, P.~Hochloff, and R.~Dargaville, ``{Flexible
  electricity tariffs: Power and energy price signals designed for a smarter
  grid},'' \emph{Energy}, vol.~93, pp. 2568--2581, dec 2015.

\bibitem{Deetjen2018}
T.~A. Deetjen, J.~S. Vitter, A.~S. Reimers, and M.~E. Webber, ``{Optimal
  dispatch and equipment sizing of a residential central utility plant for
  improving rooftop solar integration},'' \emph{Energy}, vol. 147, pp.
  1044--1059, mar 2018.

\bibitem{Ren2016}
Z.~Ren, G.~Grozev, and A.~Higgins, ``{Modelling impact of PV battery systems on
  energy consumption and bill savings of Australian houses under alternative
  tariff structures},'' \emph{Renewable Energy}, vol.~89, pp. 317--330, apr
  2016.

\bibitem{DACHpaper}
L.~Bloch, J.~Holweger, C.~Ballif, and N.~Wyrsch, ``{Impact of advanced
  electricity tariff structures on the optimal design, operation and
  profitability of a grid-connected PV system with energy storage},''
  \emph{Energy Informatics}, 2019 in press.

\bibitem{Luthander2015}
R.~Luthander, J.~Wid{\'{e}}n, D.~Nilsson, and J.~Palm, ``{Photovoltaic
  self-consumption in buildings: A review},'' \emph{Applied Energy}, vol. 142,
  pp. 80--94, mar 2015.

\bibitem{IRENA2017}
\BIBentryALTinterwordspacing
IRENA, \emph{{Electricity Storage and Renewables: Costs and Markets To 2030}},
  2017, no. October. [Online]. Available:
  \url{https://www.irena.org/-/media/Files/IRENA/Agency/Publication/2017/Oct/IRENA_Electricity_Storage_Costs_2017.pdf}
\BIBentrySTDinterwordspacing

\bibitem{Planair2019}
L.~Deschaintre and F.~Jacqmin, ``{Rapport « Observation du march{\'{e}}
  photovolta{\"{i}}que 2018 »},'' Tech. Rep., 2019.

\bibitem{stein_pvlib:_2016}
J.~S. Stein, W.~F. Holmgren, J.~Forbess, and C.~W. Hansen, ``{PVLIB}: {Open}
  source photovoltaic performance modeling functions for {Matlab} and
  {Python},'' in \emph{2016 {IEEE} 43rd {Photovoltaic} {Specialists}
  {Conference} ({PVSC})}.\hskip 1em plus 0.5em minus 0.4em\relax Portland, OR,
  USA: IEEE, 2016, pp. 3425--3430.

\bibitem{gurobi}
\BIBentryALTinterwordspacing
L.~Gurobi~Optimization, ``Gurobi optimizer 8.1, reference manual,'' 2019.
  [Online]. Available: \url{http://www.gurobi.com}
\BIBentrySTDinterwordspacing

\bibitem{pandapower.2018}
L.~Thurner, A.~Scheidler, F.~Sch{\"a}fer, J.~Menke, J.~Dollichon, F.~Meier,
  S.~Meinecke, and M.~Braun, ``pandapower — an open-source python tool for
  convenient modeling, analysis, and optimization of electric power systems,''
  \emph{IEEE Transactions on Power Systems}, vol.~33, no.~6, pp. 6510--6521,
  Nov 2018.

\end{thebibliography}
%


\end{document}